# Piloting a Full-year, Optics-based High School Course on Quantum Computing

Joel A. Walsh[1], Mic Fenech[1], Derrick L. Tucker[2], Catherine Riegle-Crumb[1], and Brian R. La Cour[3]

[1] College of Education, STEM Education, The University of Texas at Austin, Austin, TX, United States of America
[2] Office of Strategy and Policy, The University of Texas at Austin, Austin, TX, United States of America
[3] Applied Research Laboratories, The University of Texas at Austin, Austin, TX, United States of America

E-mail: joelawalsh@utexas.edu, micfenech@utexas.edu, derrick.tucker@austin.texas.edu, riegle@austin.utexas.edu, blacour@arlut.utexas.edu

## Abstract

Quantum computing was once regarded as a mere theoretical possibility, but recent advances in engineering and materials science have brought practical quantum computers closer to reality. Currently, representatives from industry, academia, and governments across the world are working to build the educational structures needed to produce the quantum workforce of the future. Less attention has been paid to growing quantum computing capacity at the high school level. This article details work at The University of Texas at Austin to develop and pilot the first full-year high school quantum computing class. Over the course of two years, researchers and practitioners involved with the project learned several pedagogical and practical lessons that can be helpful for quantum computing course design and implementation at the secondary level. In particular, we find that the use of classical optics provides a clear and accessible avenue for representing quantum states and gate operators and facilitates both learning and the transfer of knowledge to other Science, Technology, and Engineering (STEM)  skills. Furthermore, students found that exploring quantum optical phenomena prior to the introduction of mathematical models helped in the understanding and mastery of the material.

## 1. Introduction

In 1965, American engineer Gordon Moore made the observation that the number of transistors in an integrated circuit seemed to double every two years [1]. This observation has held true for decades but is now reaching an end. As the number of transistors have grown, they also have become much smaller. The silicon used in transistors imposes a lower limit on size of about seven nanometers, and quantum effects such as tunneling further limit transistor size to only a few nanometers [2]. These limitations demand a new approach to computing, one in fact that takes advantage of quantum effects.

The ideas behind quantum computing, or more generally quantum information science (QIS), began to form in the early 1980s with Paul Beniof's conception of a quantum Turing machine and Richard Feynman's keynote presentation, "Simulating Physics with a Computer" [3, 4]. Over the next few decades, a number of breakthrough quantum algorithms and communication schemes were developed, such as Shor's factoring algorithm, Grover's search algorithm, and the BB84 quantum key distribution (QKD) protocol [5, 6, 7].

Over the last four decades, quantum computers have transitioned from theory to reality. Researchers can now build and interact with working quantum computers as well as share secure private keys over thousands of miles using fielded quantum communication systems [8]. As these technologies transition from theoretical to real-world use cases, the demand for a quantum workforce has grown.

While many undergraduate institutions have formed QIS programs, the high school landscape is not as well established. The first efforts to reach into the K-12 sphere came from the Institute of Quantum Computing (IQC) at the University of Waterloo. Since 2014, the IQC has invited secondary science teachers to take part in a program called "Schrödinger's Class," where they learn pedagogical approaches to teaching concepts from quantum computing, quantum mechanics, and quantum cryptography.

In recent years, a number of new educational programs and instructional resources have become available for teaching Quantum Information Science (QIS). In 2019, researchers at FermiLab, in collaboration with local area high school teachers, developed a one-week quantum computing course designed for high school students 15 to 18 years of age [9, 10]. Later, in 2020, researchers from Virginia Tech developed a two-day outreach event for high school students on quantum mechanics, with lessons and activities based on the book *Q is for Qubit* by Terry Rudolph [11, 12]. In that same year, researchers from the Massachusetts Institute of Technology, in collaboration with The Coding School, launched the first "Qubit by Qubit" summer camp and, later, a year-long course, both of which were targeted to high school students and offered remotely.

In addition to these examples of QIS curriculum, there are many European countries where concepts from quantum mechanics like entanglement or wave-particle duality are integrated into the high school physics curriculum [13]. However, to our knowledge, the first full-year high school course on quantum computing that counted towards a student's graduation requirement was developed by us at The University of Texas at Austin (UT) and piloted at a Texas public high school in the fall of 2019. This article describes our approach and the lessons learned from the first two years piloting this course.

## 2. The school

We shall use Littlewood High School as a pseudonym for the public high school in Texas used for our pilot. The school serves a diverse community, with roughly three quarters of the student body identifying as Hispanic and over half considered economically disadvantaged. Research suggests that, too often, students from minoritized communities are not given the opportunity to engage in authentic experiences in the process of learning STEM content, and one way of increasing participation in these fields is to expose them to authentic experiences that empower them to see themselves as doers of the discipline [14]. To that end, one of the goals of our project was to facilitate the future participation of underrepresented students in STEM fields through exposure to exciting emergent technologies such as quantum computing.

Littlewood High School boasts a vibrant and unique set of programs that encourages students to pursue STEM careers. Littlewood uses an academy structure that allows students to choose various pathways depending on their interests, one of which is the STEM Academy. The STEM Academy's stated mission is to provide students with exposure to rigorous curricula and real-word applications of STEM. As a mark of distinction, students in this academy are able to earn a STEM endorsement on their high school transcript upon successful completion of one of four career pathways. During its second year of implementation (2020-2021), the quantum computing course counted as one of the STEM requirements in a STEM Academy student's pathway as a Scientific Research and Design course.

In addition to support for career pathways, Littlewood High School also has several programs that support college readiness. Littlewood was one of the first high schools in the area to join the Early College High School Initiative, which allows students to earn their Associate's degree while completing high school. In addition to being an Early College High School, Littlewood also offers several dual-enrollment courses under the UT OnRamps program that allows students the opportunity to earn college credit while still in high school. This program is currently being offered in almost 200 school districts around the state of Texas, and to date, over 100,000 students have taken these courses. Starting in the 2021-2022 academic school year, the quantum computing course will be offered as a new OnRamps course at Littlewood, and it is expected that more schools around the state will offer this course in subsequent years.

## 3. Pedagogical approach

Our course's curricular objective was to introduce background knowledge and the accompanying technical applications of quantum information science. Secondary goals were providing students with the opportunity to develop technical skills in physics, programming, cybersecurity, and mathematics as well as valuable "soft skills" in critical thinking and problem solving. With these goals in mind, our pedagogical approach can best be summarized by the following three tenets: (1) introducing quantum mechanics through light polarization, (2) conducting experiments (real or virtual) before mathematics are introduced, and (3) using "just-in-time" learning of new skills and concepts such as programming and matrices. The only prerequisite was completion of the second year of high school algebra.

Given these unique constraints, it was important to converge on an interpretation of quantum mechanics that high school students would be best equipped to understand. Following the approach of IQC, as well as previous work at UT, we focused on using the properties of light, especially polarization, to introduce quantum concepts. We made this decision based on previous student exposure, to varying degrees, to concepts such as waves, light, and trigonometry. The use of polarization was restricted to linear polarization, as concepts such as circular polarization necessitated the use of complex numbers and Euler's formula and were deemed unnecessary. While complex numbers are an important part of quantum mechanics, they are not, surprisingly, needed for most introductory topics in quantum computing. It is worth noting that, while it is possible to cover introductory quantum computing without the use of complex numbers, we did want to allow educators the flexibility to decide whether or not to include additional concepts that used them into their curriculum. To this end, we decided to place complex numbers into the curriculum for the OnRamps version of the course that will be offered starting Fall 2021.

Many introductory educational programs in quantum computing tend to focus initially on simple, mathematical abstractions and then move directly to applications. For example, a qubit may be introduced as a vector of two complex numbers and gate operations as transformations of these vectors. Since the physical concepts in quantum computing are unique and unfamiliar, we felt it important to provide students with a motivation to explain experimental results and, perhaps, glean some intuition to anticipate the accompanying mathematical formalisms. For example, students can explore the behavior of polarizers before introducing the concept of vectors and component bases.

Throughout the first year of the course, which was held in person until March 2020 due to COVID restrictions, students conducted a combination of real and virtual experiments focused on optical phenomena. Real, physical experiments eventually gave way to virtual experiments using free web-based platforms such as UT's Virtual Quantum Optics Lab (VQOL) and IBM's Quantum Composer. During the second year (2020-2021), the constraints of remote instruction led to an entirely virtual approach.

These experiments were designed to incorporate an inquiry-based approach, one that prompts students to answer scientific questions by observing and generalizing patterns. This approach has been shown to have positive outcomes on learning [15]. Optics and matrix mechanics were emphasized throughout the course as being the foundation for manipulating qubits and programming quantum algorithms. VQOL was used to simulate how light could be manipulated and represented how real quantum states could be controlled [16]. Later in the course students used IBM's Quantum Composer, a free cloud service that provides access to real and simulated quantum computers, to implement quantum protocols and algorithms.

Finally, a just-in-time teaching approach was taken to provide students with relevant skills as they became necessary, which has also been shown to have positive outcomes on learning [17]. For example, some Python programming was required for a QKD project that students completed about half-way through the year. Thus, basic Python programming skills were not taught until they were needed by students to complete the project. Students learned Python programming in Google Colab, a free cloud service that allows students to create notebooks that can execute code and rich text in a single document.

## 4. Lessons learned

As a result of both the relative novelty of high school QIS curriculum and the developing global pandemic, the research team gleaned several lessons from this mixture of quantum and K-12 science curricula. While we were aware of the possibility of imparting transferable skills and conceptual understandings, there were still some surprising developments. Over halfway through our first-year pilot, the Covid-19 pandemic shifted all instruction to a remote, online learning environment. Although a virtual classroom was not the planned setting, this condition allowed students and researchers to rethink key methods of approaching experimentation and communication.

While it is certainly expected that students would leave with a better grasp of topics central to quantum information science, like optics or waves, our end-of-the-year focus groups revealed that some of the students also left with a stronger self-reported understanding of matrix algebra and classical computing. In the case of matrix algebra, Texas state mathematics standards require that all students who take high school Algebra 2 have at least a basic exposure to matrices, although some schools choose to cover more than this. In practice, this means that these units generally focus on solving systems of linear equations using Gaussian elimination or use technology to find the reduced row echelon form of matrices that represent those systems. For nearly all of our students, this class was the first time they had seen matrices applied to any topic outside of solving linear systems. For example, all three of the students in one of the focus groups reported that they had a better grasp on how to perform matrix operations, and also remarked that they also learned how matrices can represent physical objects like optical wave plates, and that matrices can be used to model transformations. Using matrices to encode transformations of physical states is a ubiquitous concept in physics and other STEM fields and yet, surprisingly, is seldom broached in the K-12 curriculum.

Despite the fact that many of our students had previous exposure to a number of programming and computer science classes, their lack of basic understanding in information theory and classical computing was somewhat surprising. While certain topics, like binary numbers and transistors, were familiar to some students, the concept of information and the inner workings of a classical computer were very new to most. Some students had taken AP Computer Science A in Java, a course that focuses heavily on object-oriented programming. The processes that allow a computer to go from transistor to even primitive data types in a programming language are obscured completely, unless the teacher sees the need to cover them. The fact that students consistently learned about the many contrasts between classical information and quantum information throughout the class seemed to aid in their understanding. Some of the students in the focus groups commented that they gained an understanding of the physical nature of information, such as the notion that information could be represented by waves.

As previously mentioned, one of the instructional approaches taken for this course was to frontload experimentation and data analysis before mathematical formalisms were introduced. This approach to physics instruction is seen by some as a way of addressing new content standard directives involving experimentation, modeling, and uncertainty [18]. An example of this approach would be the lesson on Malus's law. In the initial pilot year, students used real lasers and polarizing filters to collect data on light intensity as they manually rotated the polarizers. (With the switch to remote learning, this experiment was conducted online.) The measured data was then exported to Desmos, a free online graphing calculator, where students could see an approximately sinusoidal curve. Students were able to determine that the graph could not be of the form $I \propto cos(\theta)$, where $I$ is the light intensity and $\theta$ is the angle between the analyzer and polarizer, as the values of $I$ are never negative. After some discussion, the students fit the curve and determined the proper relation to be $I = cos^2(\theta)$. This approach allowed students to see the noise and uncertainty inherent in experimental data. It also introduced them to the authentic experience of generalization and modeling tasks integral to many STEM fields.

Collaboration was built into nearly every activity during both the in-person and remote-learning pilot years. During the first in-person year, group engagement was generally high. During the second year, which was completely remote, students generally reported low engagement during group collaboration time, with some exceptions. (Collaborative environments usually came in the form of Zoom breakout rooms.) For groups that *did* consistently collaborate, an interesting phenomenon arose: students began to use collaboration tools well outside of what was officially prescribed. These included social applications such as Google Hangouts, Discord, and Snapchat, which were mainly used to share pictures, links, and code examples. This was no doubt due in part to the technical limitations of Zoom video conferencing. In subsequent years it may be worthwhile to facilitate student Discord servers, as they allow for the smoothest exchange of messages, live video and text chat, and code via an integrated markdown system.

## 5. Conclusion

While a quantum information-dominated future might seem distant, recent technological advances point towards the necessity for exposing STEM-focused high school students to the field. We have found that an optics-based approach provides opportunities for teachers and students alike to connect abstract quantum concepts to familiar wave phenomena. Additionally, our experience with the pilot course suggests that frontloading experimentation and data analysis will help impart an understanding of the realities of scientific work and uncertainty, and provide a basis for later understanding mathematical formalism. Our pilot suggests that our approach can provide a clear and accessible method for teaching quantum concepts that provides useful transferable STEM skills. This approach does not require specialized equipment, other than basic Internet access, and does not assume the students possess special knowledge or abilities beyond junior-level mathematics.

## Acknowledgments

This work was supported by the Office of Naval Research (ONR), under Grant No. N0014-18-1-2233.